\documentclass[twocolumn,showpacs,preprintnumbers,amsmath,amssymb,prl,superscriptaddress]{revtex4-1}
\usepackage{graphicx}
\usepackage{epstopdf}
\usepackage{dcolumn}
\usepackage{bm}
\usepackage{amsmath}
\usepackage{amssymb}
\usepackage{mathrsfs}
\usepackage{amsfonts}
\usepackage{color}

\begin{document}
%%%%%%%%%   TITLE  %%%%%%%%% 
\title{The Leggett-Garg inequality and Page-Wootters mechanism } 

%%%%%%%%% AUTHORS  %%%%%%%%% 
\author{D. Gangopadhyay}
\email{debashis@rkmvu.ac.in}
\affiliation{Department of Physics, Ramakrishna Mission Vivekananda University, Belur Matth, Howrah, West Bengal, India}

\author{A. Sinha Roy}
\email{animesh.roy@rkmvu.ac.in}
\affiliation{Department of Physics, Ramakrishna Mission Vivekananda University, Belur Matth, Howrah, West Bengal, India}

%%%%%%%%%  DATE  %%%%%%%%% 
\date{\today}

\begin{abstract}
Violation of the Leggett-Garg inequality (LGI) implies  quantum phenomena. In this light  
we establish that  the Moreva \textit{et al.} \cite{moreva} experiment demonstrating the Page-Wootter's 
mechanism \cite{wootters} falls in the quantum domain. An observer outside a 2-photons  world  does not detect
any change in the $2-$photons state,i.e. there is no time parameter for the outside observer. But an observer attached to one of the photons sees the other photon evolving and this means there is an "internal" time. The LGI is violated for the clock photon whose state evolves with the internal time as measured by the system photon.  
Conditional probabilities in this  2-photons system are computed for both sharp and unsharp measurements.
The conditional probability increases for entangled states as obtained by Page and Wootters  for  both ideal and also unsharp measurements. We discuss how the conditional probabilities can be used to distinguish between massless and massive gravitons. This is important in the context of gravitational waves. 
\end{abstract}
\maketitle
\textit{1.Introduction--}
Field theories describing gauge particles are gauge theories and are invariant with respect to some local symmetry group   transformations. In Yang-Mills theories these are the non-abelian gauge transformations corresponding to $SU(N)$ groups.
For gravity these are space-time diffeomorphisms. The invariance corresponds to local Lorentz invariance. In gauge theories different solutions arising from same initial conditions become related by the local symmetry. So the general solution of the field equations is non-unique as it contains arbitrary time-dependent functions. So a subset of initial conditions needs to be chosen. This subset is defined by what are called the lagrangian constraints. In the hamiltonian formalism this implies conditions on the allowed initial positions and momenta. Time evolution must conserve these conditions. This can lead to further constraints. A first class constraint is a dynamical quantity in a constrained hamiltonian system whose Poisson bracket with all other constraints vanishes on the constraint surface in phase space. Any constraint which is not first class is a second class constraint. Local symmetry transformations are generated by the first class constraints. So gauge theories are systems with first class constraints. 
 
An early attempt to quantise gravity was through the Wheeler-DeWitt (WD) equation
\cite{dewitt}. This results from the canonical quantisation of Einstein gravity using Dirac's constrained formalism \cite{dirac}. An embodiment of  the quantum version of the hamiltonian constraint using metric variables
gives the WD equation. But this equation is \textit{time-independent}. All observables are constant and the resulting universe is unchanging and boring.

Page and Wootters \cite{wootters} proposed that the static  universe (as observed by an \textit{external} observer) actually evolves with respect to time as seen by some \textit{internal} observer \textit{within} the universe. This is because of  quantum correlations (\textit{entanglement}) between different constituents within the universe.
Consider a direct product of two different states, each state depending on a monotonically increasing parameter as well as other variables. Then an average over this  parameter will result in a state that cannot be written as a direct product of states. \textit{This is entanglement.} Conversely, entanglement implies the existence of a monotonically increasing parameter present in the objects making up the product and this parameter is identified with  some  "internal" time.   Probabilities can be determined using these entangled states. 
Alternatively, one may first calculate the  probabilities from the direct product  of time dependent states giving time dependent probabilities . Time averaging these gives probabilities of lower values than those for the entangled states. 
So a measure of quantum correlations are the conditional probabilities described above.  

Entanglement of  states is  a  non-classical feature of the quantum world. Here Bell's inequality (BI)\cite{bell} provides  a tool to study  "quantumness" or  specifically "non-locality". BI sets a bound on a certain combination of correlation functions corresponding to results of measurements on two spatially separated systems.  Suitable relative orientations of these measurements exist for which BI is violated by the relevant quantum mechanical (QM) results   for appropriate states of the entangled systems. Numerous experiments \cite{aspect} have proved the empirical violation  of BI, consistent with the QM predictions. Subsequently the Leggett-Garg inequality (LGI) \cite{leggett, leggett1} was discovered. This is a temporal analogue of BI in terms of time-separated correlation functions corresponding to successive measurement outcomes for a temporally evolving system. While furnishing a signature of distinctly quantum behaviour, LGI complements BI in providing  insight into physical reality manifested by non-classicality of quantum systems. LGI has been used  for probing possible limits of quantum mechanics in the macroscopic regime in various scenarios \cite{gangopadhyay,form,van,kofler,ruskov,gossin,laloy, souza, waldherr, athalye,rastegin,nikitin,chen}.  
For particular mention, reference \cite{form} demonstrates LGI violations over large macroscopic distances.  

Consider a two-states system which oscillates between the states $1$ and $2$ in time. Let $Q(t)$ be an observable taking values $\pm 1$ whenever measured, depending on whether the system is in state $1$ or $2$. Now consider a collection 
of runs starting from identical initial conditions such that on the first series of runs $Q$ is measured at times 
$t_{1},t_{2}$, on the second at $t_{2},t_{3}$, on the third at $t_{3},t_{4}$, and on the fourth at $t_{1},t_{4}$ with 
$t_{1}<t_{2}<t_{3}<t_{4}$. The expression $[Q(t_{1})Q(t_{2}) + Q(t_{2})Q(t_{3}) + Q(t_{3})Q(t_{4}) - Q(t_{1})Q(t_{4})]$ is always $+2$ or $-2$. 
The temporal correlations $C_{ij} \equiv \langle Q(t_{i})Q(t_{j})\rangle$ are determined. Replacing individual product terms  by their averages over the entire ensemble of such sets of runs, the LGI is \cite{leggett, leggett1} :
\begin{equation}
\label{1}
C \equiv C_{12} + C_{23} + C_{34} - C_{14} \leq 2
\end{equation} 
This is a  a Bell-type inequality. The $t_{i}$'s play the role of apparatus settings.  
This inequality imposes realistic constraints on time-separated joint probabilities pertaining to oscillations in any two-states system. Violation of this inequality signifies quantum phenomena. 

Sometime back Moreva \textit{et al.} \cite{moreva} in a remarkable experiment illustrated the Page and Wooters' mechanism of "static" time (paragraph 3)  by using an entangled state of the polarisation 
of two  photons. One was used as a clock to measure the evolution of the other as follows: an "internal" observer that became correlated with the clock photon sees the other system evolve, while an "external" observer recording only global properties of the two photons sees no change. The motivation of our present work is to consider this experiment in the light of the LGI (Section 2). We also seek further evidence for the Page-Wootters mechanism by computing relevant conditional probabilities for both ideal(Section 3) and unsharp (Section 4) measurements. Finally, we discuss how the conditional probabilities may provide a route to distinguish between massive and massless gravitons (Section 5). Concluding remarks are in Section 6.

\textit{2.The Page-Wootters Mechanism--}
Page and Wootters used  a static entangled state $|\psi\rangle$ whose subsystems evolve according to quantum mechanics for an observer who uses one of the subsystems as a clock system $c$ to measure the time evolution of the rest $r$. Subsystems are assumed to be non-interacting. The hamiltonian of the \textit{global} system is written as  $\mathscr{H} =\mathscr{H}_{c}\otimes 1_{r}+1_{c}\otimes \mathscr{H}_{r}$. Here $\mathscr{H}_{c}, \mathscr{H}_{r}$ are \textit{local} hamiltonians  for $c$ and $r$, respectively \cite{wootters}. The state of the "universe" $|\psi\rangle$ is then determined by enforcing the WD equation $\mathscr{H} |\psi\rangle =0$. So $\psi$ is an eigenstate of $\mathscr{H}$ with eigenvalue zero.  The reason for this choice is that by projecting $\psi$ on the states $|\phi (t)\rangle _{c}=e^{-i\mathscr {H}_{c}t/\hbar}|\phi (0)\rangle _{c}$ of the clock, one gets  $|\phi (t)\rangle _{r}$ \textit{which is defined to be} $|\psi (t)\rangle _{r}:=\,_{c}\langle \phi (t)|\psi\rangle=e^{-i\mathscr {H}_{r}t/\hbar}|\psi (0)\rangle _{r}$. This describes a evolution of the subsystem $r$ under its local hamiltonian $\mathscr{H}_{r}$, the initial state being $|\psi (0)\rangle _{r}:=\,_{c}\langle \phi (0)|\psi\rangle$ . Hence, although \textit{globally} the system appears static, its components (i.e.\textit{locally}) exhibit correlations that signify dynamical evolution \cite{wootters}.

In \cite{moreva} an entangled quantum state $|\psi\rangle$ of two photons , clock photon $c$ with horizontal polarisation $|H\rangle$ and system photon $r$ with vertical polarisation $|V\rangle$, is represented as
\begin{equation}
\label{2}
|\psi\rangle=\frac{1}{\sqrt{2}}(|H\rangle _{c}|V\rangle_{r}-|V\rangle_{c}|H\rangle_{r})
\end{equation}
The hamiltonian of the global system is $\mathscr{H} =\mathscr{H}_{c}\otimes 1_{r}+1_{c}\otimes \mathscr{H}_{r}$ with 
$\mathscr{H}_{c}=\mathscr{H}_{r}=i\hbar\omega (|H\rangle\langle V|-|V\rangle\langle H|)$. 
Here $\mathscr{H}|\psi\rangle =0$ i.e. the \textit{$2-$photons state} satisfies the WD equation \cite{dewitt}. So an observer external to  the $2-$photons world does not detect any change in the $2-$photons state. 
So there is no time parameter for the outside observer. But an observer attached to one of the photons (i.e. internal observer) sees the \textit{single photon state} corresponding to the other photon evolving and this signifies existence
of an "internal" time. Here we study the quantum mechanical violation of LGI for the clock photon whose state evolves with the internal time as measured by the system photon.

Let  the clock photon be in state $|H\rangle$ at $t=0$ i.e., $|\phi (0)\rangle _{c}=|H\rangle$. Denote this initial condition by $1$. The time evolved state of clock photon is
\begin{eqnarray}
\label{3}
&&|\phi (t)\rangle _{c}=e^{-i\mathscr {H}_{c}t/\hbar}|\phi (0)\rangle _{c}= e^{-i\mathscr {H}_{c}t/\hbar}|H\rangle\nonumber\\
&&=\Big(1-\frac{i\mathscr{H}_{c}t}{\hbar}
+\cdots\Big)|H\rangle = \cos (\omega t)|H\rangle -\sin (\omega t)|V\rangle
\end{eqnarray}
So after time $t$ the probability of getting  horizontal (vertical) polarization $|H\rangle$  ($|V\rangle$) are
$P_{1H}(t)=\cos ^{2}(\omega t)~~;~~ P_{1V}(t)=\sin ^{2}(\omega t)$.
If at time $t=0$ the clock photon is in state $|V\rangle$, initial condition $2$,
the time evolution will be 
\begin{eqnarray}
&&exp(-iH_{c}t/\hbar)|V\rangle =\cos (\omega t)|V\rangle +\sin (\omega t)|R\rangle
\end{eqnarray}
so that now 
$P_{2V}(t)=\cos ^{2}(\omega t)~~;~~P_{2H}(t)=\sin ^{2}(\omega t)$
where the suffixes $1,2$ denote the initial conditions.
Now start an experiment with the clock photon in state $|H\rangle$ at $t=0$. Then the joint probability 
$P_{1H,H}(t_{1},t_{2})$ of  finding the state $|H\rangle$ at $t_{1}$ and again $|H\rangle$ at $t_{2}$ 
(where $0<t_{1}<t_{2}$) is 
\begin{eqnarray}
P_{1H,H}(t_{1},t_{2}) =\cos ^{2}(\omega t_{1})\cos ^{2}\{\omega (t_{2}-t_{1})\}
\end{eqnarray}
Similarly one can calculate $P_{1H,V}(t_{1},t_{2})$, $P_{2V,H}(t_{1},t_{2})$ and $P_{2V,V}(t_{1},t_{2})$. Now consider an observable quantity $Q(t)$ such that, whenever measured, it takes values  $+1$ or $-1$ depending on whether the system is in $|H\rangle$ or $|V\rangle$ state. Then the time correlation function $C_{12}=\langle Q(t_{1})Q(t_{2})\rangle$ can be evaluated by using above mentioned four joint probabilities to obtain
\begin{eqnarray}
\label{8}
C_{12}=\cos ^{2}\{\omega (t_{2}-t_{1})\}-\sin ^{2}\{\omega (t_{2}-t_{1})\}
\end{eqnarray}
$C_{12}$ depends on the temporal separation $(t_{2}-t_{1})$ and the parameter $\omega$ (which defines the time scale of the system). In general 
$C_{ij}=\cos ^{2}\{\omega (t_{j}-t_{i})\}-\sin ^{2}\{\omega (t_{j}-t_{i})\} ,~~ j > i$. Using these equations 
\begin{eqnarray}
\label{9}
&&C\nonumber\\
&&=\cos ^{2}\{\omega (t_{2}-t_{1})\}+\cos ^{2}\{\omega (t_{3}-t_{2})\}+\cos ^{2}\{\omega (t_{4}-t_{3})\}\nonumber\\
&&+\sin ^{2}\{\omega (t_{4}-t_{1})\}-\cos ^{2}\{\omega (t_{4}-t_{1})\}-\sin ^{2}\{\omega (t_{2}-t_{1})\}\nonumber\\
&&-\sin ^{2}\{\omega (t_{3}-t_{2})\}-\sin ^{2}\{\omega (t_{4}-t_{3})\}\nonumber\\
&&=\cos ^{2}(x)+\sin ^{2}(3x)-3\sin ^{2}(x)-\cos ^{2}(3x)
\end{eqnarray}
where we have chosen $(t_{4}-t_{3})=(t_{3}-t_{2})=(t_{2}-t_{1})=\Delta t$, and defined $x=\omega \Delta t$. The behaviour of the quantity $C$ with $x$ is shown in the figure 1.
\begin{figure}
\resizebox{7.0cm}{4.5cm}{\includegraphics{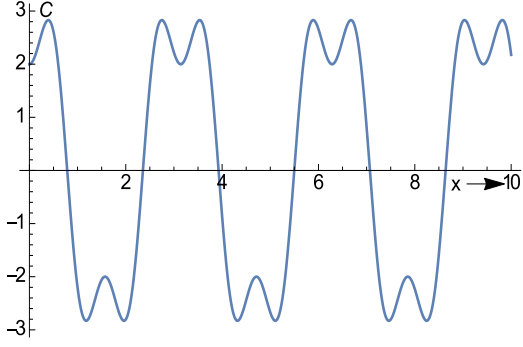}}
\caption{Variation of $C$ against $x=\omega\Delta t$}
\end{figure}
$C$ oscillates with time. The maximum upper bound for $C$ is $C_{max}=2\sqrt{2}=2.82843$.

Therefore  the results  of this section establishes the fact that the same photon at different times is auto-correlated and the Moreva \textit{et al} \citep{moreva} experiment falls in the quantum domain with maximum 
violation of the LGI for the clock photon whose state evolves with the internal time as measured by the system photon.

\textit{3.Conditional Probabilities for sharp measurements--}
We now compute the conditional probabilities following Page and Wootters \cite{wootters}. We first discuss sharp (i.e. projective or ideal) measurements. At time $t=0$, let the clock photon be in state $|H\rangle$ and  the system photon be in state $|V\rangle$. After time $t$ the clock photon state is $\cos \omega t|H\rangle _{c}-\sin \omega t|V\rangle _{c}$ and system photon state becomes   $\cos \omega t|V\rangle _{r}+\sin \omega t|H\rangle _{r}$. The time averaged (stationary) state 
$|\bar\psi \rangle$ of the two photon system is then
\begin{eqnarray}
\label{11}
|\bar\psi \rangle \propto && \int_{0}^{2\pi/\omega} \big(\cos \omega t|H\rangle _{c}-\sin \omega t|V\rangle _{c}\big)\times \big(\cos \omega t|V\rangle _{r}\nonumber\\
&&-\sin \omega t|H\rangle _{r}\big)dt\nonumber\\
&& \propto |H\rangle _{c} |V\rangle _{r} - |V\rangle _{c} |H\rangle _{r}
\end{eqnarray}
Hence the normalized state of the two particle system is 
\begin{equation}
\label{12}
|\bar\psi \rangle = \frac{1}{\sqrt{2}}\big( |H\rangle _{c} |V\rangle _{r} - |V\rangle _{c} |H\rangle _{r}\big)
\end{equation}
If the clock photon is found in the horizontal polarization, then the probability of finding system photon with vertical polarization will be 
\begin{equation}
\label{13}
P\Big(|V\rangle_{r} \Big\vert |H\rangle_{c}\Big) 
= \frac{\vert (\,_{c}\langle H|_{r}\langle V|)|\bar\psi\rangle \vert ^{2}}{\vert \,_{c}\langle H|\bar\psi\rangle \vert ^{2}} =1
\end{equation} 
Now consider the time dependent $2-$photons state 
\begin{eqnarray}
\label{14}
|\psi (t) \rangle = (\cos \omega t|H\rangle _{c}-\sin \omega t|V\rangle _{c})(\cos \omega t|V\rangle _{r}+\sin \omega t|H\rangle _{r})\nonumber\\
\end{eqnarray}
With this state the \textit {the time averaged  probability} for system photon to be vertically polarised is
\begin{eqnarray}
\label{15}
\bar{P}\Big(|V\rangle_{r} \Big\vert |H\rangle_{c}\Big) =
 \frac{\int_{0}^{2\pi/\omega} \vert (_{c}\langle H|_{r}\langle V|)|\psi (t)\rangle\vert ^{2}dt}
{\int_{0}^{2\pi/\omega} \vert \,_{c}\langle H|\psi (t)\rangle\vert ^{2} dt}
 = \frac{3}{4}
\end{eqnarray}
The above two conditional probabilities are next determined using density matrix formalism.
Consider the stationary state $|\bar{\psi} \rangle$ (\ref{12}). The density matrix corresponding to this state is
\begin{eqnarray}
\label{16}
\rho_{\bar\psi} = \frac{1}{2}[|H\rangle_{c}|V\rangle_{r}-|V\rangle_{c}|H\rangle_{r}][\,_{c}\langle H|_{r}\langle V|-\,_{c}\langle V|_{r}\langle H|]\nonumber\\
\end{eqnarray}
With  this $\rho_{\bar\psi}$ (clock photon horizontally polarized), probability of finding vertically 
polarized system photon is
\begin{eqnarray}
\label{17}
P_{\rho_{\bar\psi}}\Big(|V\rangle _{r} \Big\vert |H\rangle_{c}\Big)
 =\frac{Tr[\mathscr{P}_{HV} \rho_{\bar{\psi}}]}{Tr[\mathscr{P}_{H} \rho_{\bar{\psi}}]}
=1
\end{eqnarray}
$\mathscr{P}_{H_{c}V_{r}}=|H\rangle_{c}|V\rangle_{r} \,_{c}\langle H|_{r}\langle V|$ is the projection operator where clock photon is in state $|H\rangle _{c}$ while system photon is 
in state $|V\rangle _{r}$. $\mathscr{P}_{H_{c}}=|H\rangle _{c} \,_{c}\langle H|$ is the projection operator for clock photon in state $|H\rangle _{c}$ .

Next consider the time dependent state (\ref{14}). Here the density operator $\rho_{\psi} =|\psi (t) \rangle\langle \psi (t)|$ contains 16 terms, and
$Tr[P_{H_{c}V_{r}} \rho_{\psi}]=\cos ^{4} \omega t$ and $Tr[P_{H_{c}} \rho_{\psi}]=\cos ^{2} \omega t $.
So when clock photon is horizontally polarized, the time-averaged conditional probability of finding vertically polarised  system photon is
\begin{eqnarray}
\label{20}
\bar{P}_{\rho_{\psi}}\Big(|V\rangle _{r}\Big\vert |H\rangle_{c}\Big) = \frac{\int_{0}^{2\pi/\omega} Tr[P_{HV} \rho_{\psi}] dt}{\int_{0}^{2\pi/\omega} Tr[P_{H} \rho_{\psi}]dt} =\frac{3}{4}
\end{eqnarray}
So (\ref{13}),(\ref{15}), (\ref{17}), (\ref{20}) show that for pure states there are no differences between the conditional probabilities calculated using the density matrices or otherwise. This is as it should be. 

\textit{4.Conditional Probabilities for unsharp measurements--}
Now consider unsharp (or non-projective, non-ideal) measurements \cite{busch}. For a sharp measurement of the polarisation of a single (say $i-$th ) photon the dichotomic observable is (no sum over $i$)
$Q_{i}=|H\rangle_{i} \,_{i}\langle H|-|V\rangle _{i} \,_{i}\langle V|=P_{i+}-P_{i-}$, with outcomes $\pm 1$ for photon in state $|H\rangle$ or $|V\rangle$ respectively. Corresponding projection operators are 
$\mathscr{P}_{i\pm} =\frac{1}{2}(I_{i}\pm Q_{i})$ where $I_{i}=|H\rangle_{i}\,_{i}\langle H|+|V\rangle _{i}\,_{i}\langle V|$. For \textit{unsharp measurements} a sharpness parameter $\lambda$ is introduced to characterize the precision of a measurement and $\lambda=1$ means projective ("sharp") measurement. The unsharp projection operators are now  
\begin{equation}
\label{21}
\mathscr{F}_{i\pm} = \frac{1}{2}(I_{i}\pm \lambda _{i} Q _{i}) = \lambda _{i} \mathscr{P}_{i\pm} +(1-\lambda _{i})\frac{I _{i}}{2}
\end{equation}
$(0 <\lambda _{i} < 1)$, $\mathscr{F}_{i\pm}$ are mutually commuting operators with non-negative eigenvalues and 
$\mathscr{F}_{i+}+\mathscr{F}_{i-}=I$.
$\lambda _{i} =1$ corresponds to sharp measurements and  $\mathscr{F}_{i\pm}$ reduce to $\mathscr{P}_{i\pm}$. For a sharp measurement of the polarisation of two photons , the projection operator for  clock  photon  in $|H\rangle$ state and system photon in $|V\rangle$ state is
\begin{eqnarray}
\label{22}
\mathscr{P}_{H_{c}V_{r}}=\frac{1}{4}(I_{c} + Q_{c})(I_{r} - Q_{r}) = |H\rangle_{c}|V\rangle_{r} \,_{c}\langle H|_{r}\langle V| \nonumber\\
\end{eqnarray}
For the unsharp measurement this operator becomes
\begin{eqnarray}
\label{23}
&&\mathscr{F}_{H_{c}V_{r}}=\frac{1}{4}(I_{c} + \lambda Q_{c})(I_{r} -\lambda Q_{r})\nonumber\\
&& = \frac{1}{4}\Big[ (1+\lambda _{c})(1-\lambda _{r})|H\rangle_{c}|H\rangle_{r} \,_{c}\langle H|_{r}\langle H|+(1+\lambda _{c})\nonumber\\
&&(1+\lambda _{r})|H\rangle_{c}|V\rangle_{r} \,_{c}\langle H|_{r}\langle V|+(1-\lambda _{c})(1-\lambda _{r})|V\rangle_{c}|H\rangle_{r} \nonumber\\
&& _{c}\langle V|_{r}\langle H| +(1-\lambda _{c})(1+\lambda _{r})|V\rangle_{c}|V\rangle_{r} \,_{c}\langle V|_{r}\langle V| \Big]
\end{eqnarray}
So for the unsharp measurement on the state (\ref{16}) (clock photon with horizontal polarization), the probability of finding the system photon with vertical polarization is
\begin{eqnarray}
\label{24}
P_{\rho _{\bar{\psi}}}\Big(|V\rangle _{r} \Big\vert |H\rangle_{c}\Big) =\frac{Tr[\mathscr{F}_{H_{c}V_{r}} \rho _{\bar{\psi}}]}{Tr[\mathscr{F}_{H_{c}} \rho _{\bar{\psi}}]} =\frac{1}{2}(1+\lambda _{c} \lambda _{r})
\end{eqnarray}
Now  consider unsharp measurements for \textit{time dependent} $2-$photons state with density matrix $\rho_{\psi} =|\psi (t) \rangle\langle \psi (t)|$ where $\psi (t)$ is given by equation (\ref{14}) . Here
\begin{eqnarray}
\label{25}
Tr[\mathscr{F}_{H_{c}V_{r}} \rho _{\psi}]&&=\frac{1}{4} \Big[(1+\lambda _{c})(1-\lambda _{r}) \sin ^{2} \omega t \cos^{2} \omega t\nonumber\\
&& (1+\lambda _{c})(1+\lambda _{r})\cos^{4}\omega t+(1-\lambda _{c})(1-\lambda _{r})\nonumber\\
&& \sin^{4} \omega t+(1-\lambda _{c})(1+\lambda _{r})\sin^{2} \omega t \cos^{2} \omega t\Big] \nonumber\\
\end{eqnarray}
and
\begin{eqnarray}
\label{26}
Tr[\mathscr{F}_{H_{c}} \rho _{\psi}]=\frac{1}{2} \Big[1+\lambda _{c}\cos 2\omega t\Big]
\end{eqnarray}
So now if clock photon is horizontally polarized, probability of vertically polarised  system photon is 
\begin{eqnarray}
\label{27}
\bar{P}_{\rho _{\psi}}\Big(|V\rangle _{r} \Big\vert |H\rangle_{c}\Big) = \frac{\int_{0}^{2\pi/\omega} Tr[F_{H_{c}V_{r}} \rho _{\psi}] dt}{\int_{0}^{2\pi/\omega} Tr[F_{H_{c}} \rho _{\psi}]dt} =\frac{1}{4}\Big[2+\lambda _{c}\lambda _{r}\Big]\nonumber\\
\end{eqnarray}

\begin{figure}
\resizebox{7.0cm}{4.5cm}{\includegraphics{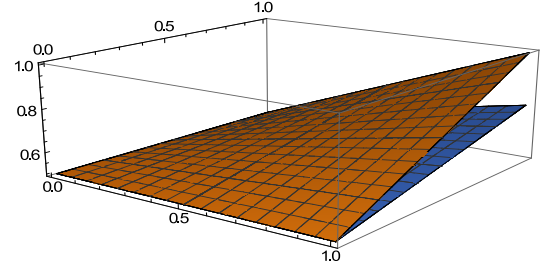}}
\caption{3D-Plot of conditional probabilities with $\lambda _{c} , \lambda _{r}$. Horizontal axes 
represent $\lambda _{c}$ ,$\lambda _{r}$. Vertical axes represent values of the two probabilities}
\end{figure}

\begin{figure}
\resizebox{7.0cm}{4.5cm}{\includegraphics{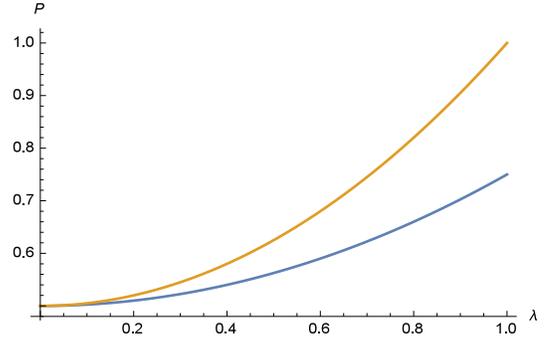}}
\caption{2D-Plot of conditional probabilities with $\lambda _{c}= \lambda _{r}=\lambda $}
\end{figure}
For $\lambda _{c} = \lambda _{r} =1$, (\ref{24})  and (\ref{27}) for unsharp measurements  match exactly with the values of ideal measurements i.e. $1$ and $\frac{3}{4}$ respectively. Figure 2 gives the variation of the two conditional probabilities for unsharp measurements (\ref{24}) and (\ref{27}) where $0\leq (\lambda _{c} , \lambda _{r})\leq 1$. The upper graph is for two entangled photons while the lower graph is for two un-entangled (i.e. time dependent states) photons. The upper curve is for entangled photons while the lower curve is for time dependent photon wavefunctions. \textit{The probability for entangled (i.e. time independent state) photons is always greater than that of the un-entangled photons for all non-zero values of $\lambda _{c} , \lambda _{r}$}. Figure 3 is for $\lambda _{c} = \lambda _{r} =\lambda$. . 

However, an important point must be remembered. Page and Wootters considered \textit{massive} spin particles. A \textit{massive} spin $j$ particle will have $2j+1$ spin angular momentum eigenstates. We are considering photons which are \textit{massless} gauge particles. The gauge constraint allows only two independent degrees of freedom i.e. two polarisations. So our calculated probabilities correspond to two-state systems . In fact,  our calculated probabilities are  identical to that for electrons which are also two state systems (j= $1/2$ and $2j+1=2$). Note that our result is different from their values for \textit{massive}  spin one particles because of the above reasons. 
In Section 5 we discuss how this can help distinguish between massless and massive gravitons.

\textit{5. Conditional probabilities and the graviton--}
Counting degrees of freedom for a \textit{massless} graviton goes as follows. The \textit{massless} graviton is the quantum of the gravitational field whose \textit{classical} limit is general relativity. Here the metric tensor 
$g_{\mu\nu}= g_{\nu\mu}$ is a symmetric tensor with 16 components. For a symmetric tensor in $4-$dimensions the number of independent components are  $\frac{12}{2}=6$ (off-diagonal components)\textit{plus} $4$ (diagonal components)$=10$. The Bianchi identities reduces the number of degrees of freedom by $4$. Invariance under space-time diffeomorphisms (i.e. invariance under general coordinate transformations) removes another $4$ unphysical degrees of freedom. So finally only two degrees of freedom are left. So in our formalism, the massless graviton is very similar 
to that of the photon. Generalising to $D$ dimensions, a massless graviton has $\frac{D(D-3)}{2}$ 
degrees of freedom (d.o.f.) in D-dimensional spacetime. A massive graviton has $\frac{D(D-1)}{2}-1$ d.o.f. in D-dimensional spacetime. So in $4$ dimensions this is $5$. Therefore, it is like a massive particle with spin $j=2$, where the number of spin states will be $ 2j + 1= 5$. Hence the conditional probabilities would be similar to those of a massive spin $2$ particle. \textit {What matters is the number of degrees of freedom and not the refined details 
of the specific theories describing the particle}. So by determining the relevant conditional probabilities one should be able to distinguish between massless and massive gravitons. 

\textit{6.Concluding remarks--} 
We first consider the Moreva \textit{et al} \citep{moreva} experiment in the light of the LGI (Section 2). Recall that an "external" observer will see no dynamics of the two photons , i.e. "static" universe. However, an observer attached to one of the photons will see the other evolve if the photons are entangled. So 
\textit{temporal evolution$\Longleftrightarrow$ entanglement/correlations}. Each of the two photons (i.e."clock photon" and "system photon") individually violates the maximum bound of LGI and this implies that this is a quantum phenomenon. We  show that autocorrelations between the \textit{same photon at different times} are present as revealed by the maximal violation of the LGI. The time variable of the LGI is the \textit{internal time as seen by the other photon}. So applicability of the LGI means existence of an internal time as seen by the other photon.  
We next compute the relevant conditional probabilities for ideal and non-ideal measurements (Sections 3 and 4)in the 2-photon system (photons are massless spin one particles) and show that the conditional probability increases for entangled states as obtained by Page and Wootters  for  both ideal and also unsharp measurements. Section 5 discusses how the relevant conditional probabilities may help in distinguishing between massless and massive gravitons. 

\textit{7.Acknowledgements-}
A. Sinha Roy's work was done under UGC-CSIR Research Fellowship No. 2061151173.

\end{document}